\documentclass[aps,preprint,prb,superscriptaddress]{revtex4}
\usepackage{amsfonts}
\usepackage[final,hiresbb]{graphicx}
\usepackage{amsmath}
\usepackage{amssymb}
\usepackage{amstext}

\usepackage{color}
\usepackage{braket}

\usepackage{subfigure}

\definecolor{Green}{RGB}{0,204,102}
\definecolor{Purple}{RGB}{102,0,255}
\definecolor{Blue}{RGB}{0,0,255}
\definecolor{Red}{RGB}{151,010,010}

\begin{document}

\title{Twisted Molecular Excitons as Mediators for Changing the Angular Momentum of Light}  
\author{Xiaoning Zang and Mark T. Lusk}
\email{mlusk@mines.edu}
\affiliation{Department of Physics, Colorado School of Mines, Golden, CO 80401, USA}

\begin{abstract}
Molecules with $C_N$ or $C_{Nh}$ symmetry can absorb quanta of optical angular momentum to generate \emph{twisted excitons} with well-defined quasi-angular momenta of their own. Angular momentum is conserved in such interactions at the level of a paraxial approximation for the light beam. A sequence of absorption events can thus be used to create a range of excitonic angular momenta.  Subsequent decay can produce radiation with a single angular momentum equal to that accumulated. Such molecules can thus be viewed as mediators for changing the angular momentum of light. This sidesteps the need to exploit nonlinear light-matter interactions based on higher-order susceptibilities. A tight-binding paradigm is used to verify angular momentum conservation and demonstrate how it can be exploited to change the angular momentum of light.  The approach is then extended to a time-dependent density functional theory setting where the key results are shown to hold in a many-body, multi-level setting.
\end{abstract}
\maketitle

\section{Introduction}

Angular momentum (AM) about a fixed point is conserved in isolated systems characterized by a rotationally symmetric Lagrangian~\cite{Noether_1971}. The eigenstates of atomic electrons exemplify this, and the AM of the collective light-matter system is  conserved in association with their spontaneous orbital decay~\cite{Grinter_2008}. For systems with less symmetry, though, there is no expectation that a one-to-one correspondence should exist between the electronic and photonic units of AM in light-matter interactions~\cite{Andrews_2010_1, Andrews_2010_2}. 

Of course individual photons, as bosonic field excitations, have a well-defined helicity of unit magnitude~\cite{Leader_2016}. In addition, electromagnetic radiation fields described by the time-ordered superpositions of such excitations can also have a photonic angular momentum (PAM) even when full rotational symmetry is not present. A Gaussian beam of circularly polarized light is composed of photons of the same helicity and has a PAM of $^\pm 1$ in units of $\hbar$. Within the paraxial approximation, optical vortices~\cite{Allen_1992} also have a quantized AM per photon that are eigenvalues of an operator that does not depend on gauge or frame~\cite{QED_BLP_1982, Barnett_2016, Leader_2016}.  A circularly polarized Laguerre-Gaussian beam of azimuthal index $l$, for instance, has PAM = $^\pm 1+l$ within the paraxial approximation. These AM measures are possible because the radiation has an axis of symmetry and a negligibly small radial gradient.

\begin{figure}[hptb]
\begin{center}
\includegraphics[width=0.9\textwidth]{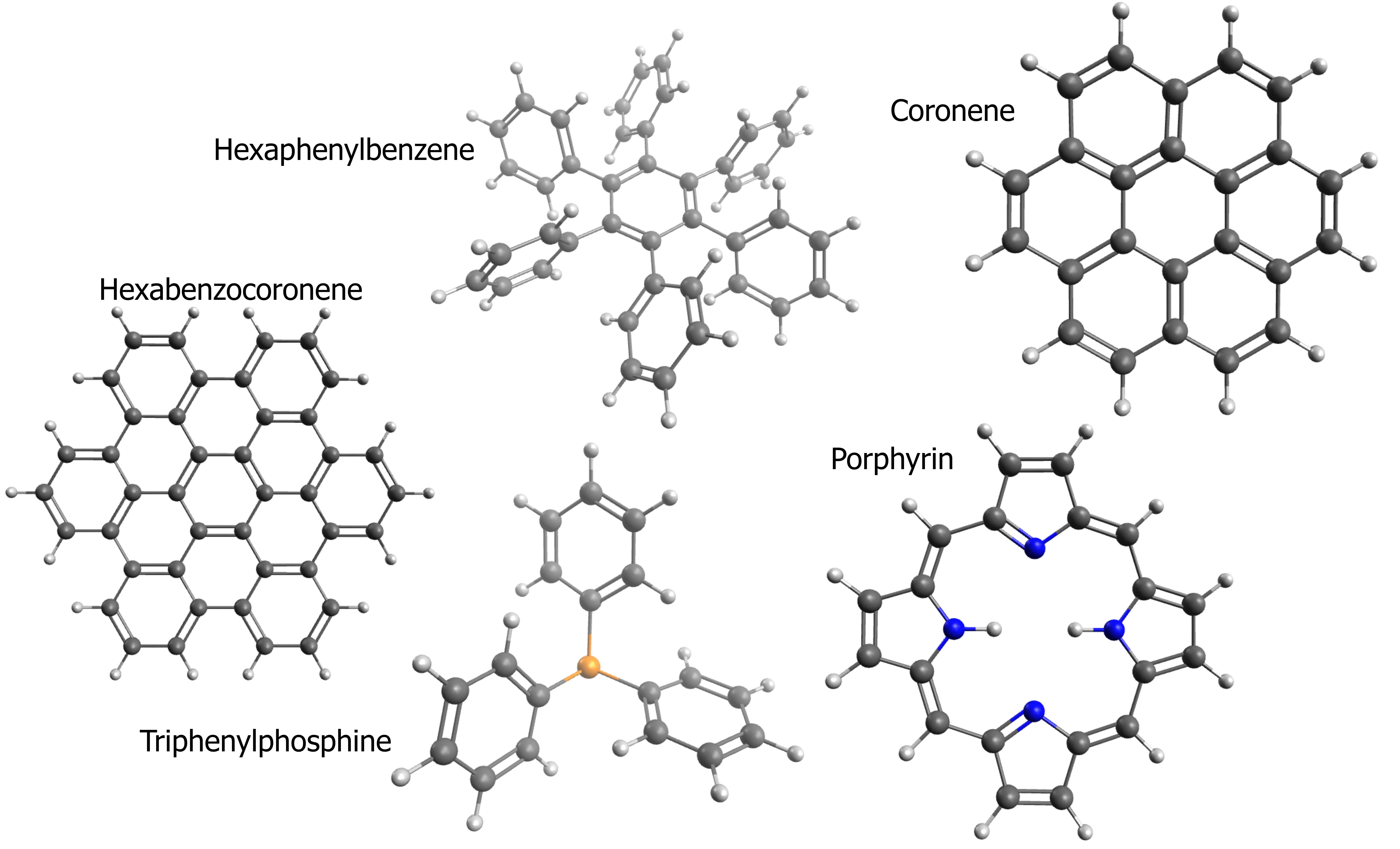}
\end{center}
\caption{\emph{Molecules with $C_N$ or $C_{Nh}$ symmetry}. The three planar molecules shown are among many polycyclic aromatic hydrocarbons for which there exists conjugation between adjacent arms. Three-dimensional molecules, such as triphenylphosphine and hexaphenylbenzene, may also offer sufficiently strong coupling between arms.} 
\label{molecules}
\end{figure}

A molecular axis of symmetry facilitates the assignment of a meaningful AM to the electronic state of specially designed molecules as well.  These have a discrete rotational symmetry group of either $C_N$ or $C_{Nh}$, such as those shown in Figure \ref{molecules}, and their repeated subunits are referred to as arms. These may radiate outward as chiral spokes, as in triphenylphosphine ($\rm{Ph}_3\rm{P}$) and hexaphenylbenzene, or may compose a planar arrangement as found in porphyrin and corronene structures.
%
%

The associated molecular eigenstates can be expressed as a phase-shifted superposition of the lowest energy excitons associated with each of the arms. These collective states are identifiable by the phase shift, $2 \pi q_e/N$, between neighboring arms in the superpositions~\cite{AndrewsPRL2013, Andrews_2014}. The integer, $q_e$, is just the number of $2\pi$ windings of the phase accumulated in traversing the circuit of arms and so can be interpreted as  the number of units of an excitonic, quasi-angular momentum.

Quasi-momenta are encountered in any material system that exhibits a discrete symmetry. For instance, solid-state lattices have a quasi-linear momentum that is distinct from the particle momentum of the ions and electrons of which it is composed. This momentum is the conserved quantity associated with a discrete translational symmetry of the lattice~\cite{Ashcroft_1976}. An analogous application of discrete rotational invariance has been used to explain how particle angular momentum can be transferred to a quasi-angular momentum for a variety of quasi-particles~\cite{Baltz_2002}. When the lattice points fall on a circle, as with the molecular arms of centrosymmetric molecules, this quasi-angular momentum is simply the product of an effective radius, $R$, with excitonic quasi-momentum, $\hbar k_{ex}$. Azimuthal periodicity then implies that $k_{ex} R$ is an integer, $q_e$. In analogy to their photonic counterparts~\cite{Padgett_2004, Dennis_2009}, these electronic states are referred to here as \emph{twisted excitons} with an excitonic angular momentum (EAM) of $q_e$. Each molecular eigenstate has a distinct EAM, and there is a one-to-one relationship between the eigenenergies and the magnitude of EAM.

\begin{figure}[hptb]
\begin{center}
\includegraphics[width=0.7\textwidth]{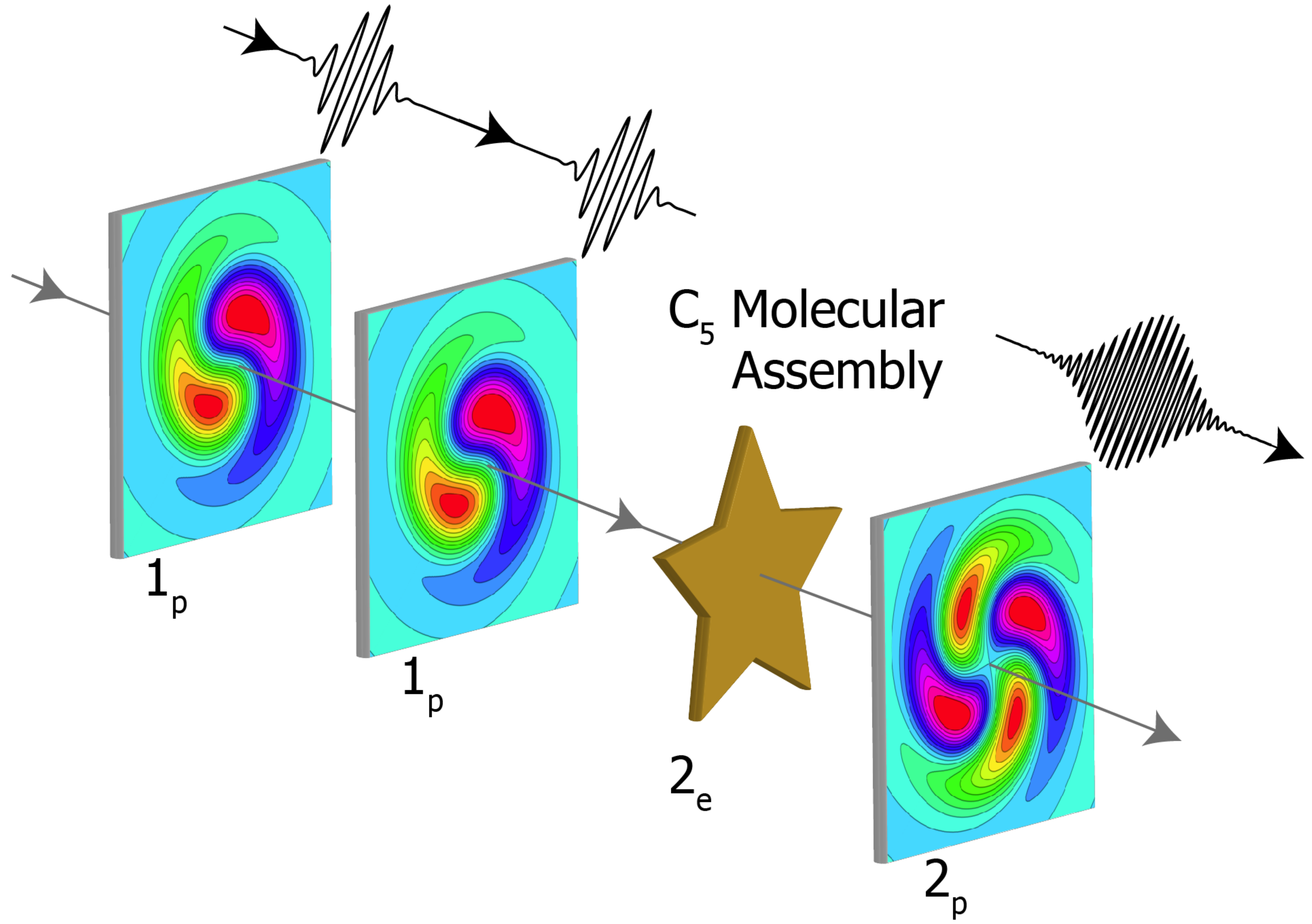}
\end{center}
\caption{\emph{Changing the angular momentum of light using twisted excitons.} Two laser pulses, each with a PAM = $1_p$, are sequentially absorbed by a molecular assembly resulting in an EAM = $2_e$. The radiation subsequently emitted has a PAM = $2_p$.}\label{Calculator}
\end{figure}

Electronic decay of these molecules results in the generation of an optical vortex\cite{AndrewsPRL2013} in which both energy and AM are transferred between excitonic and optical forms. The same is true for absorption events in which this process is reversed and, in fact, for a sequence of such absorption events.  Subsequent emission, either spontaneous or stimulated, can produce radiation with the accumulated EAM. In this way, the excitons play the role of an angular momentum bank in which PAM can be deposited and withdrawn in different increments. The concept, illustrated in Figure \ref{Calculator}, offers a means of changing the AM of light that does not rely on higher-order susceptibilities~\cite{SHGPRA1996, SHGPRA1997,FWMPRL2012, FWMOL2016,FWMOE2009, FEMPRB2015, FWMPRB2016}. Angular momentum is conserved in these light-matter interactions, and a one-to-one correspondence between excitonic and photonic angular momentum therefore exists in this special setting.

%
%

This particular form of AM conservation is in contrast to that associated with the excitation or decay of a single electron state for hydrogen-like atoms lying on the axis of an optical vortex. Suppose that a static magnetic field makes absorption/emission preferred in one direction, and consider a right-circularly polarized Laguerre-Gaussian beam, within the paraxial approximation, of azimuthal index $l = 1$ that is aligned with it. The electromagnetic radiation therefore has PAM = $2_p$. Even though the electric field intensity is zero along the symmetry axis, it has a nonzero gradient, and the light-matter quadrupole coupling, $M = \bf{Q}\cdot \nabla \bf{E}$, is finite. Here $\bf{Q}$ is the transition quadrupole tensor between the $1s$ and $3d_2$ states. Two units of AM can therefore be transferred from beam to electron as has only recently been experimentally confirmed~\cite{Schmiegelow_2016}. Here the electrons have a continuous azimuthal symmetry as opposed to the discrete excitonic symmetry of centrosymmetric molecules.

Electromagnetic radiation composed of a superposition of photons of both helicities exhibits AM conservation in each of its components. A second single-electron example makes this clear, and the idea extends to the molecular setting as well.  The same Laguerre-Gaussian beam of azimuthal index $l = 1$, but now linearly polarized, interacts with the atomic ground state, and the electron can once again be raised to the $3d_2$ state. This would seem to violate AM conservation because the aggregate input beam has PAM = $1_p$ and results in an EAM = $2_e$ state. However, the beam is actually composed of a superposition of right and left circularly polarized optical vortices that have AM of $2$ and $0$, respectively. The AM of each individual interaction is conserved: $1_{ps}+1_{pl} \rightarrow 2_e$ and $^-1_{ps}+1_{pl} \rightarrow 0_p$. Here subscripts $p$ and $e$ indicate photonic and excitonic manifestations while subscript $s$ and $l$ denote polarization and vortex origins.  For such mixed state beams, electronic occupation probability must be included in the AM algebra: $1_p \rightarrow \frac{1}{2}2_e + \frac{1}{2}(^-1_{ps} + 1_{pl})$.

We examine AM banking in association with centrosymmetric molecules using a combination of theory and computation. A tight binding (TB) paradigm is used to analytically prove that AM is conserved in association with sequences of absorption and emission. Numerical implementations demonstrate this within a time-dependent setting. The associated Hamiltonian is then replaced with one that does not rely on prescribed transition dipoles and for which electron excitations are treated as many-body events with exchange energies and correlation effects included. Twisted excitons are no longer just superpositions of two-level excitations on each molecular subunit, but AM conservation emerges none-the-less. This time-domain Density Functional Theory (TD-DFT) setting is used to simulate the dynamics of AM transfer between photonic and excitonic manifestations, and simple additions and subtractions are once again demonstrated. In both TB and TD-DFT settings, the underlying processes are linear in the sense that only first-order electric dipole interactions are necessary. This is in contrast to the nonlinear optics strategy for up/down converting angular momentum using higher order susceptibilities. There is a tradeoff, certainly, because the AM banking scheme presumes that exciton relaxations and decoherence processes occur on a time scale slower than that of the AM manipulations. For the sake of clarity in exploring this alternative methodology, though, exciton-phonon coupling and dynamic disorder are disregarded.   

\section{Tight-Binding Paradigm}\label{TB}
 
First consider a molecular Hamiltonian in the absence of light-matter coupling. The requisite $C_N$ or $C_{Nh}$ symmetry is provided by an N-arm molecule in which arm $j$ supports two energy levels: ground state $\ket{\xi^{j,0}}$ and excited state $\ket{\xi^{j,1}}$. The tight-binding (TB) Hamiltonian is taken to be
\begin{equation}
\hat{H}_0=\sum_{j=1}^{N}\Delta \hat{c}_j^{\dagger}\hat{c}_j+\sum_{<i,j>}^{N}(\tau\hat{c}_j^{\dagger}\hat{c}_i+H.c.)
\label{TBH0}
\end{equation} 
Here $\Delta$ is the excited state energy of each arm, $\tau$ is the coupling between nearest arms, and $\hat{c}_j^{\dagger}$ is the creation operator for arm $j$. It is straightforward to show~\cite{AndrewsPRL2013} that the ground state is $\ket{0}=\prod_{j=1}^{N}\ket{\xi^{j,0}}$ while the N excited states are
\begin{equation}
\ket{v_{q_e}}=\sum_{j=1}^{N}\frac{\varepsilon^{(j-1)q_e}}{\sqrt{N}}\ket{e_j}
\label{TBstates}
\end{equation}
with $\varepsilon=\mathrm{e}^{\imath 2\pi/N}$ and  $\ket{e_j}=\ket{\xi^{j,1}}\prod_{m\neq j}^{N}\ket{\xi^{m,0}}$. The EAM, $q_e$, is an integer with values bounded by $\frac{-1}{2}(N-1)$ and $\frac{1}{2}(N-1)$. The  corresponding energies are 
\begin{equation}
\mathbb{E}_{q_e}=\Delta + 2\tau \cos\biggl(\frac{2\pi q_e}{N}\biggr),
\label{TBenergies}
\end{equation}
with $p=q_e$ for $q_e\ge 0$ and $p=q_e+N$ for $q_e<0$. A hollow $\mathbb{E}$ will be used to distinguish exciton energy from electric field, $E$.

Now introduce semi-classical light-matter coupling via two Hamiltonians: $\hat H_1$ that governs light-mediated interactions between the ground state and each molecular eigenstate, while $\hat H_2$ governs the analogous laser interactions that mediate transitions between eigenstates.  The angular momentum of incident electric fields may be manifested as a circular polarization, a vector vortex, or linear polarization with a scalar vortex, but we restrict attention to the first two types. An electric dipole approximation is made, and the discrete rotational symmetry ensures that a rotation of the molecule about its axis by $2\pi/N$ maps one dipole into the next. Under these conditions, the details of electric field structure and dipole orientations are irrelevant, and the light-matter interactions are well-captured by the following Hamiltonians, which are functions of the PAM of the incident light, $q_p$:
\begin{eqnarray}
\hat H_1(q_p)&=& -\mu_{0}^* E  \sum_j \varepsilon^{-q_p(j-1)} \hat{c}_j^{\dagger}\hat{c}_0 + H.c. \nonumber \\
\hat H_2(q_p) &=& -\mu_{{\rm hop}}^* E \sum_j \varepsilon^{-q_p(j-1)} \hat{c}_{{\rm mod}(j)+1}^{\dagger}\hat{c}_j + H.c.
\label{H12}
\end{eqnarray}
The \emph{mod} function returns its argument modulo $N$ and use has been made of the fact that $\varepsilon^{-q_p({\rm mod}(j)-1)} = \varepsilon^{-q_p(j-1)}$. The scalars, $\mu^*_0 E$ and $\mu_{{\rm hop}}^* E$, represent the inner product of electric transition dipole moments with a time-dependent electric field.

The total Hamiltonian, $\hat H(q_p) = \hat H_0 + \hat H_1(q_p) + \hat H_2(q_p)$, is then applied to the Schr{\" o}dinger equation with solutions assumed to be of the form
\begin{equation}
\ket{\Psi(t)} = A_0(t) \ket{0} + \sum_{n=1}^N A_n(t) \ket{e_n}.
\label{LCAO}
\end{equation}
This results in a set of $N+1$ coupled ODE's that can be solved numerically for a prescribed electric field and initial state. The evolving state can then be projected onto each excitonic eigenstate to determine the population of each AM, $q_e$, as a function of time:
\begin{equation}
\rho_{q_e}(t) := |\braket{\Psi(t)|v(q_e)}|^2 = \biggl(\sum_{n=1}^{N}  A^*_n(t) \frac{\varepsilon^{(n-1)q_e}}{\sqrt{N}}\biggr)^2 .
\label{pop}
\end{equation}
%

\subsection{Conservation of Angular Momentum}
Suppose that the molecule is initially in eigenstate $\ket{v_{I_e}}$, where subscript $I_e$ indicates an initial EAM of $I$. A beam with PAM = $q_p$ is incident on the molecule, exciting it into eigenstate $\ket{v_{F_e}}$. Subscripts $p$ and $e$ delineate photonic and excitonic manifestations while $F$ is the EAM of the final state. A necessary condition for this transition to occur is that $H_{IF} := \bra{v_{F_e} }\hat H \ket{v_{I_e}}\ne 0$. Assuming that $I_e\ne F_e$, Equations \ref{TBstates} and \ref{H12} imply that
\begin{widetext}
\begin{equation}
H_{IF} = -\mu_{\rm hop}^* E \biggl(\sum_i \frac{\varepsilon^{-(i-1)F_e}}{\sqrt{N}}\bra{e_i}\biggr)  \biggl(\sum_j \varepsilon^{-(j-1)q_p} \hat{c}_{{\rm mod}(j)+1}^{\dagger}\hat{c}_j\biggr)   \biggl(\sum_k \frac{\varepsilon^{(k-1)I_e}}{\sqrt{N}}\ket{e_k}\biggr),
\end{equation}
which can be easily reduced to
\begin{equation}
H_{IF} =  -\frac{\mu_{\rm hop}^* E \varepsilon^{-F_e}}{N}  \biggl(\sum_j \varepsilon^{-(j-1)(I_e - q_p - F_e)}\biggl) - \frac{\mu_{\rm hop}^* E \varepsilon^{I_e}}{N}  \biggl(\sum_j \varepsilon^{-(j-1)(I_e + q_p - F_e)}\biggl) .
\label{twoterms}
\end{equation}
\end{widetext}
The following  cyclic sum orthogonality property of periodic exponentials is then useful:
\begin{equation}
\sum_j \varepsilon^{-(m-n)j} = N \delta_{m,n}.
\end{equation}
It is applied to both terms in Equation \ref{twoterms} to give
\begin{equation}
H_{IF} =  -\mu_{\rm hop}^* E \varepsilon^{-F_e}  \delta_{I_e, q_p + F_e} - \mu_{\rm hop}^* E \varepsilon^{I_e} \delta_{I_e + q_p, F_e} .
\end{equation}

Resonant illumination implies that exactly one of the two Kronecker delta functions will be nonzero. We therefore have the following statements of AM conservation:
\begin{eqnarray}
\mathbb{E}_{I_e} &<& \mathbb{E}_{F_e} \quad \rightarrow \quad I_e + q_p = F_e \nonumber \\
\mathbb{E}_{I_e} &>& \mathbb{E}_{F_e} \quad \rightarrow \quad I_e  = F_e + q_p .
\label{conservation}
\end{eqnarray}
If the initial exciton energy is lower than that of the final state, absorption results in an increase in the quasi-angular momentum of the molecule. The converse is also true in association with radiation. Since the sign of the PAM can be either positive or negative, this allows for a number of ways in which the electromagnetic field can be used to manipulate the level of excitonic angular momentum. It also offers a strategy for withdrawing angular momentum from the molecule in a range of denominations. This is next illustrated in two applications.  

\subsection{Tight-Binding Implementations of Angular Momentum Conservation}

As a first proof-of-concept, a 7-arm molecule with an EAM = $2_e$ is subjected to windowed, continuous wave (CW) lasers with a PAM = $^-1_p$ and the following scalar waveform:
\begin{equation}
E(t) = E_0 \,{\rm Sin}\bigl((\mathbb{E}_{F_e} -\mathbb{E}_{I_e})t/\hbar\bigr) .
\label{TB_E}
\end{equation}
As usual, $I_e$ and $F_e$ are the initial and final AM of the exciton with corresponding energies, $\mathbb{E}_{I_e}$ and $\mathbb{E}_{F_e}$, given by Equation \ref{TBenergies}. The parameters used, and the resulting molecular eigenstate energies and EAM values, are given in Table \ref{TB_data}.

%
\begin{table}[hptb]
\caption{Exciton energies and AM associated with 7-arm molecule used for all TB simulations. The following parameters were used (atomic units): $E_0 = 2\times 10^{-4}, \mu_0=1, \Delta= 1, \tau = 0.067$ and a radial distance to the center of each arm of 0.6. Here $\mu_0\equiv\mu_{{\rm hop}}$ is the strength of the transition dipole. Proportional changes to these parameters do not affect the results.\label{TB_data}}
\begin{ruledtabular}
\begin{tabular}{lccccccr}
EAM& $^-3_e$ & $3_e$ & $^-2_e$ & $2_e$ & $^-1_e$ & $1_e$ & $0_e$  \\ 
Energy (Ha) & 0.880 & 0.880 & 0.970 & 0.970 & 1.083 & 1.083 & 1.133\\ 
\end{tabular}
\end{ruledtabular}
\end{table}

\begin{figure}[hptb]
\begin{center}
\includegraphics[width=0.7\textwidth]{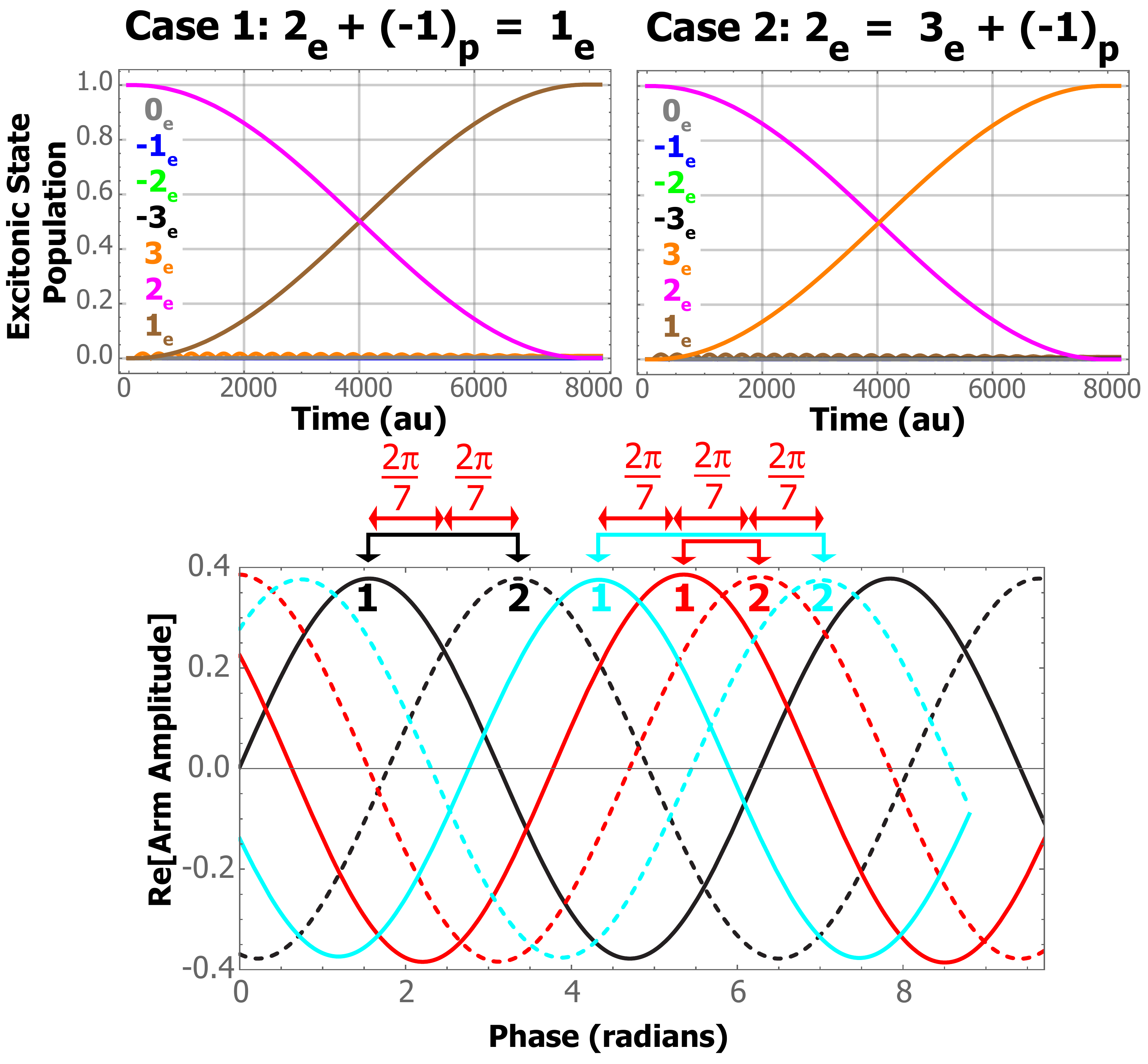}
\end{center}
\caption{\emph{Addition and subtraction of EAM on a 7-arm molecule.} (a, left): Laser energy is the difference between the second and first eigenstates and results in the absorption of radiation. (a, right): Laser energy is the difference between the second and third eigenstates and results in the emission of radiation. In both cases the laser is applied throughout the entire simulation shown. All parameters are listed in Table \ref{TB_data}. Panel (b) shows phase progression between neighboring arms for both cases as detailed in the text. The color legends of the top panels identify the populations of each EAM state.}
\label{TB_AM2_up_down}
\end{figure}

The top plots of Figure \ref{TB_AM2_up_down} show how the AM can be transferred from a laser to the molecule. Two scenarios are considered, and the associated AM conservations are listed above each plot. In both, the molecule has an initial EAM = $2_e$, and the associated exciton energy is 0.970 Ha. A laser of the form of Equation \ref{TB_E} is applied. In the left plot, the laser has a PAM = $^-1_p$, and its frequency is set to difference between the energies of states for which EAM = $2_e$ and EAM = $1_e$---i.e. $0.113/\hbar$ Ha. Energy conservation then prevents any meaningful increases in the population of states other than those for which EAM = $^\pm 1_e$. However, angular momentum conservation prevents the growth of the EAM = $^-1_e$ state. The result is that radiation is absorbed, the exciton energy is increased to 1.083 Ha, and the EAM is reduced to a value of $1$.  It is worth emphasizing that energy increases are associated with EAM decreases.(See Table \ref{TB_data}.) 

%
%
%

The top-right plot in Figure \ref{TB_AM2_up_down} demonstrates how a change to the laser frequency can cause radiation to be emitted instead. Here the frequency is set to difference between the energies of states for which EAM = $2_e$ and EAM = $3_e$---i.e. $0.09/\hbar$ Ha. As in the left plot, the laser PAM is set to $^-1_p$. Energy conservation then allows the growth of only the EAM = $^\pm 3_e$ states, but angular momentum conservation prevents the growth in population of the state for which EAM = $^-3_e$. The result is that radiation is emitted, the exciton energy is decreased to 0.880 Ha, and the EAM is increased to a value of $3_e$. These examples demonstrate that a combination of laser frequency and AM is sufficient to either add or subtract AM from the exciton.

The bottom plot in Figure \ref{TB_AM2_up_down} is a composite of results from both simulations which shows the phase relationship between two adjacent arms before and after application of the laser pulses. The horizontal axis was obtained by rigidly translating sections of the plots of amplitude versus time so that they appear in the same time interval. Then time was mapped into phase through multiplication with the associated frequencies of light. In this form, the phase relation between two arms can be easily measured by comparing the amplitude of one arm (solid) with its neighbor (dashed). The initial phase progression should be $4 \pi/7$, from Equation \ref{TBstates}, and this is confirmed in the black and dashed black curves. The addition of PAM results in an EAM = $1_e$ and the measured phase between the red and dashed red curves exhibits the expected progression of $2\pi/7$. Likewise, the subtraction of PAM leaves the molecule with EAM = $3_e$ and the anticipated phase progression of $6\pi/7$ between arms, shown in blue and dashed blue.

\begin{figure}[hptb]
\begin{center}
\includegraphics[width=0.7\textwidth]{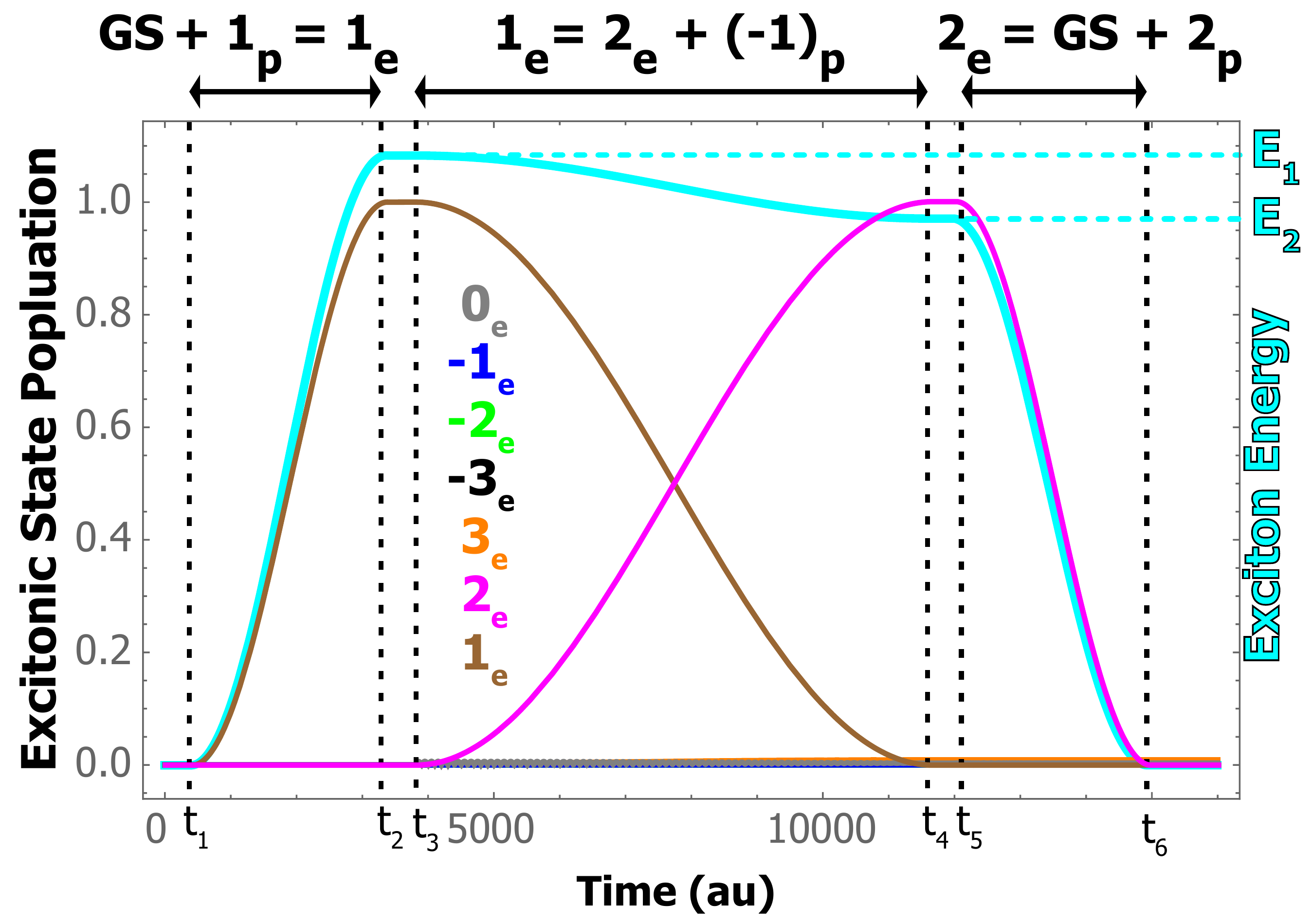}
\end{center}
\caption{\emph{PAM doubling.} 7-arm molecule initially in ground state (GS) is subjected to a sequence of 3 lasers. The excitonic state populations evolve in accordance with conservation of AM as noted at the top of the plot. Two $1_p$ pulses are absorbed and one $2_p$ radiation results. The color legend identifies the populations of each EAM state. Note that the dark green curve and dashed lines are associated with the exciton energy with scale at right.}
\label{TB_addition}
\end{figure}

A second TB simulation, Figure \ref{TB_addition}, carries out a sequence of AM addition and subtraction that starts and ends in the ground state (GS). The same 7-arm molecule (Table \ref{TB_data}) is subjected to three windowed, CW laser pulses. The first laser $(t_1 \le t_2)$ has PAM = $1_p$ and an energy equal to that of the eigenstate for an EAM $= 1_e$. The second laser $(t_3 \le t_4)$ has the same PAM but with an energy equal to the difference of excitonic states associated with EAM = $1_e$ and $2_e$. The PAM of the third laser is $^-2_p$ with an energy equal to that of the eigenstate for which EAM = $2_e$. The AM balances associated with each laser are given in the figure to more easily interpret the data plotted. These plots also show how the excitonic energy evolves as a function of time, becoming asymptotic to the appropriate eigenenergies after each laser pulse is applied. The simulation shows that a sequence of absorption events can be followed by a single emission with the latter having a PAM equal to the sum of the input PAMs.

%
%
%

\section{Angular Momentum Transfers with Time-Domain Density Functional Theory}\label{TD-DFT-Results}

Many of the idealizations associated with the TB paradigm can be removed by reconsidering the light-matter dynamics using time-domain Density Functional Theory (TD-DFT). Unlike standard ground state Density Functional Theory (DFT), TD-DFT captures the non-equilibrium response of material to an externally applied, time-varying electric field. Such real-time simulations are made possible through Runge-Gross (RG) reformulation of the time-dependent Schr{\" o}dinger equation~\cite{RGtddftPRL1984}.  A methodology was developed so that TD-DFT can be used to quantify AM transfers as detailed in the Methods section.

TD-DFT calculations are computationally intense and amount to carrying out a standard DFT calculation for a series of very small time steps. The requisite time step for the calculations of this study is $0.027$ a.u. for simulations covering approximately $2067$ a.u. in total. To reduce the computational cost in this initial proof-of-concept, a ring of radially aligned $H_2$ molecules was used as an idealized N-arm system. The ring was given a radius of $3.8$ Bohr with the H-H bond lengths taken to be $1.4$ Bohr. For each simulation, a computational domain was constructed as the sum of spheres of radius $5.7$ Bohr around each atom. This domain was discretized with a spacing of $0.28$ Bohr. The generalized gradient approximation (GGA) parametrized by Perdew, Burke, and Ernzerhof (PBE)\cite{PBE} was adopted to account for exchange and correlation, and a Troullier-Martins pseudopotential was used.  Since the wavelength of the requisite laser field is much larger than the dimension of the N-arm system, a point dipole approximation of light-matter interaction is applied in our TD-DFT simulations.

Limitations on the type of external field that can be inputted to the TD-DFT routine necessitated a piecewise construction to approximate  optical vortices. Transition dipoles from the ground state were found to be maximal on each arm and so the associated N field components were constructed so as to be arm-centered.  In contrast, the transition dipoles between two excited states were maximal at the midpoints between arms, so the associated external field components were centered between the arms. 

A 5-arm configuration of $H_2$ dimers was adopted in all TD-DFT simulations. Casida's perturbative TD-DFT methodology~\cite{casida1995response} was first performed to obtain excitation energies. This was necessary in order to design laser pulses with the frequency needed to excite a given excitonic state. The energies, dominant determinants, and the corresponding EAM of the first five excited states are given in Table~\ref{CasidaExs}. The determinants listed there are those that dominate each excited state, representing approximately $90\%$ of the respective wavefunction. Approximating an excited state with only a single determinant makes it possible to find the population of EAM states, Equation~\ref{population}, as detailed in the Methods section.

%
\begin{table}[hptb]
\caption{Dominant determinant, $\Psi_a^r$, and EAM, $q_e$, of the first five lowest excited states as calculated from Casida perturbation within TD-DFT. Here $\Psi_a^r$ is the spin-adapted singlet so that one electron is excited from $a^{th}$ occupied KS orbital to the $r^{th}$ unoccupied KS orbital. Energies are in Hartrees (Ha). \label{CasidaExs}}
\begin{ruledtabular}
\begin{tabular}{lccccr}
Excited State& $1$ & $2$ & $3$ & $4$ & $5$  \\ \hline
Energy (Ha) & 0.37 & 0.37 & 0.39 & 0.39 &0.41\\
Determinant & $\Psi_5^6$ & $\Psi_4^6$ & $\Psi_3^6$ & $\Psi_2^6$ & $\Psi_1^6$\\
Percents (\%) & 97 & 97 & 94 & 94 & 91\\
$q_e$ &\multicolumn{2}{c}{$^\pm2$} &\multicolumn{2}{c}{$^\pm1$} &0\\
\end{tabular}
\end{ruledtabular}
\end{table}

\begin{figure}[hptb]
\begin{center}
\includegraphics[width=0.9\textwidth]{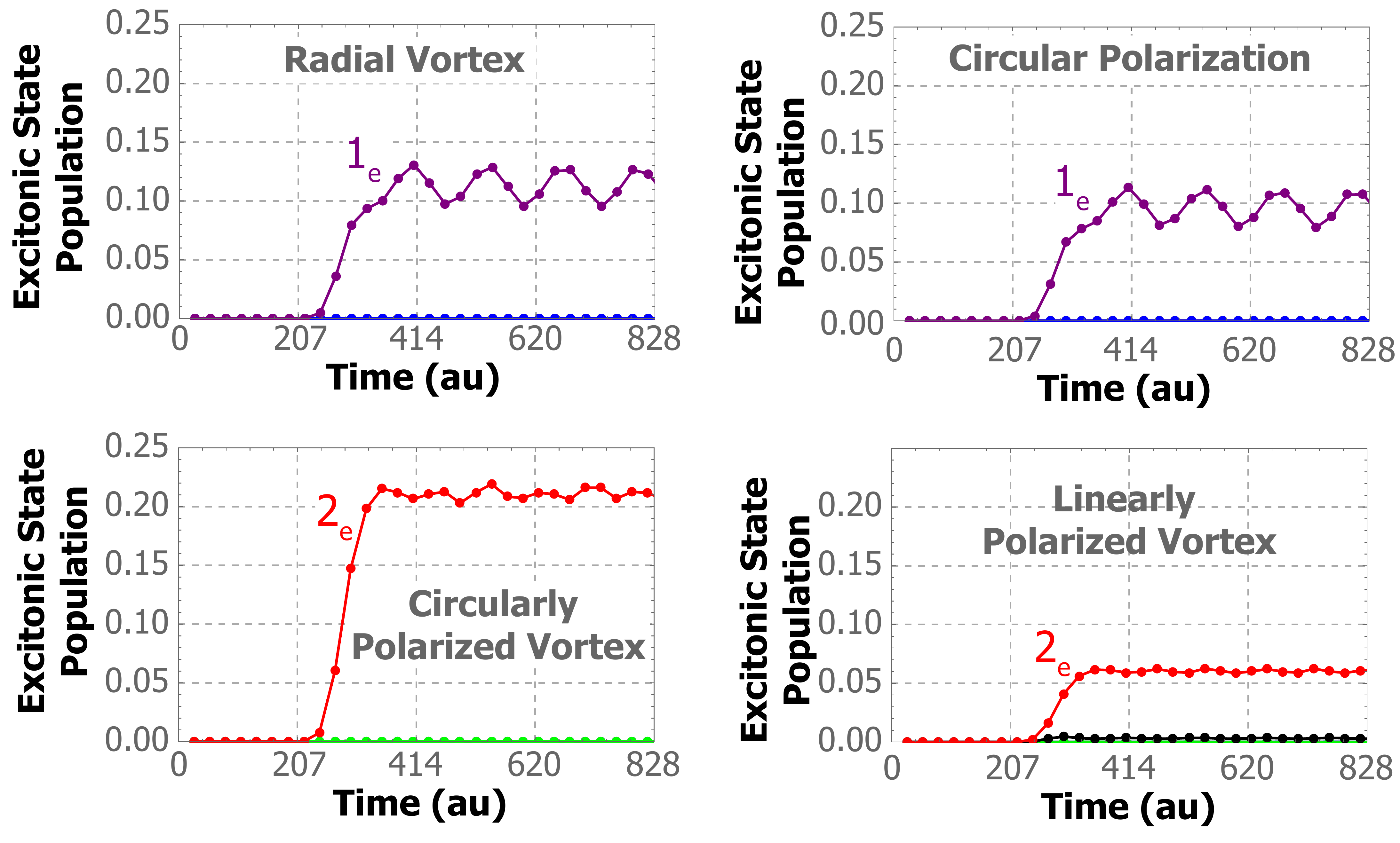}
\end{center}
\caption{\emph{TD-DFT excitations of 5-arm centrosymmetric system.} The plots show the evolution of excitonic state populations in response to: (a) radial vector vortex with energy 0.39 Ha and AM $^-1_{l}$; (b) circularly polarized light with energy 0.39 Ha and AM $^-1_{s}$; (c) circularly polarized vortex with energy 0.37 Ha and combined AM = $^-1_{l} + ^-1_{s}$; and, (d) a linearly polarized vortex with energy $0.37$ Ha and AM = $^-1_{l}$. The Gaussian envelope, $E_0\, \mathrm{e}^{-(t-t_0)^2/2\tau^2}$, has parameters $\{t_0 = 272, \tau=27.2, E_0 = 0.0100\}$ in atomic units.}\label{ExcitationTDDFT}
\end{figure}

A radial vector vortex, carrying PAM = $^-1_{pl}$ and energy of 0.37 Ha excites the 5-arm system from its ground state to the EAM = $^-1_e$ state as shown in Figure~\ref{ExcitationTDDFT} (a). Panel (b) of the figure shows that the same state can be achieved using circularly polarized light, $^-1_{s}$. Combining these two photonic structures and using a radiation energy of 0.39 Ha results in an EAM =$^-2_e$ state, as expected (panel (c)). On the other hand, the linearly polarized vortex of PAM = $^-1_{l}$ results in the same EAM =$^-2_e$ state but with only one-half the occupation probability (panel (d)). This is because the beam can be decomposed into vortices with opposing circularly polarizations, one with a combined PAM of $^-1_{s}$ + $^-1_{l}$ and the other with a PAM of $^+1_{s}$ + $^-1_{l}$.

%
%
%

\subsection{Algebra of Radial Vector Vortex Fields}

A radial vortex was chosen to test conservation of AM for the more realistic many-body TD-DFT Hamiltonian. Three cases were considered which each involve a sequence of two laser pulses. The incident radiation field, with a Gaussian envelope, was taken to be: 
\begin{equation}
\vec{E}(t) = E_0\mathrm{e}^{-\mathrm{i}q_e\phi}\mathrm{e}^{\mathrm{i\omega t}}\mathrm{e}^{-(t-t_0)^2/2\tau^2} \vec{e}_r .
\label{gaussian}
\end{equation}
As with the TB analyses, conservation of energy determines whether or not PAM is added to or subtracted from the molecular assembly. 

\begin{figure}[hptb]
\begin{center}
\includegraphics[width=0.9\textwidth]{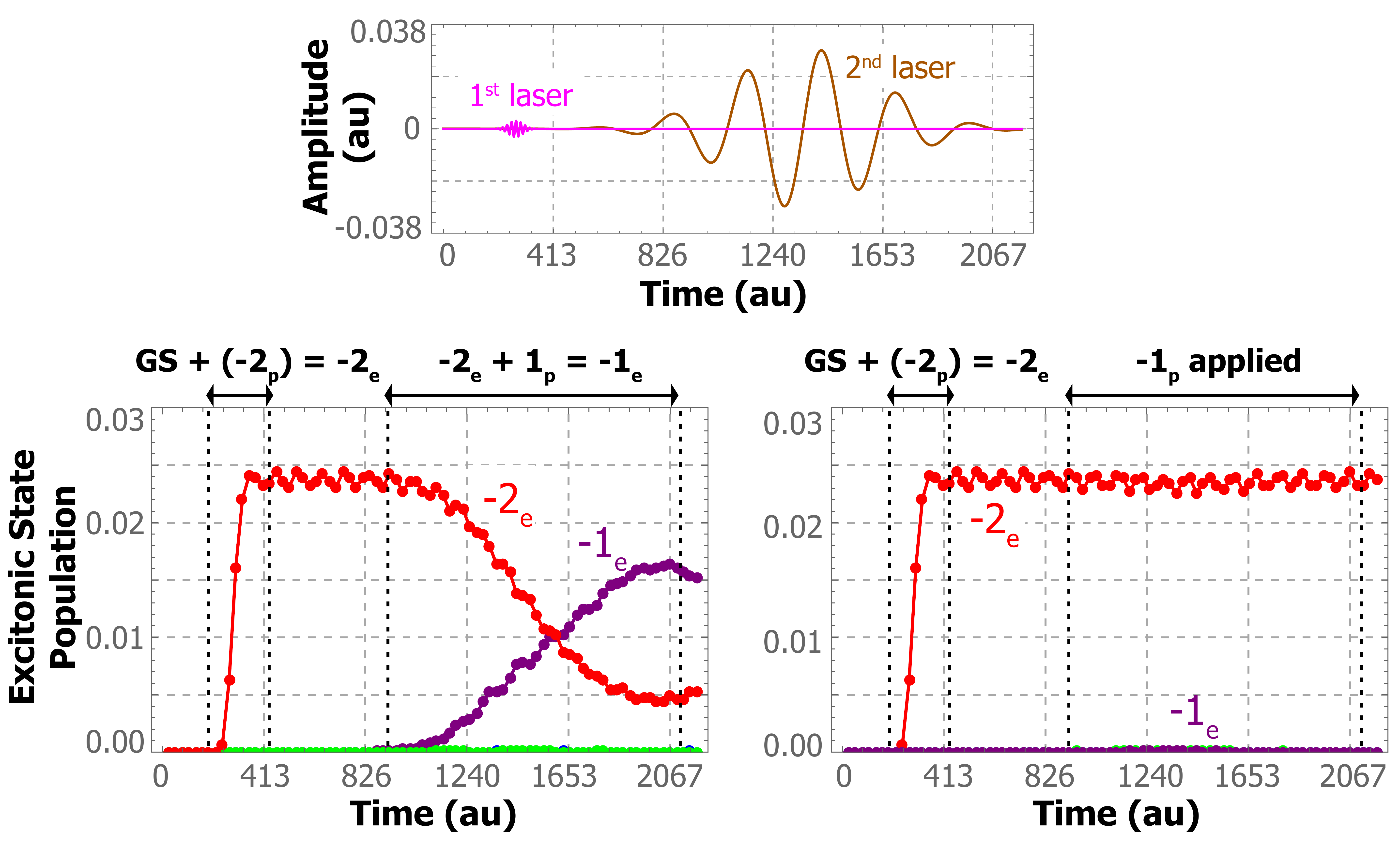}
\end{center}
\caption{\emph{TD-DFT AM manipulation:} $^-2_e + 1_p$ = $^-1_e$. Panel (a) shows sequenced laser pulses of 0.367 Ha and 0.0220 Ha, respectively. The system is initially given an EAM = $^-2_e$ using a $2_p$ laser. Application of second laser with AM = $1_p$ transfers the system to a $^-1_e$ state as shown at left in panel (b). The same laser energy, but with an AM = $^-1_p$, does not cause the state to evolve as shown at right in panel (b). Envelop parameters for Equation \ref{gaussian} are $\{t_0 = 272, \tau=27.2, \omega=0.367, E_0 = 0.00300\}$ and $\{t_0 = 1360, \tau=272, \omega=0.0220, E_0 = 0.0300\}$ in atomic units, respectively. }\label{ETCm2tom1}
\end{figure}

A first comparative analysis demonstrates how conservation of AM can be used to control exciton manipulations.  The sequence of radiation absorption events shown at bottom-left in Figure~\ref{ETCm2tom1}. A 0.367 Ha laser pulse with PAM = $^-2_p$ is first used to create an exciton with EAM = $^-2_e$. This has the highest magnitude of quasi-angular momentum possible and so is of the lowest energy, as listed in Table~\ref{CasidaExs}.  A second laser pulse, with an energy equal to the difference between that of the first and second excitonic states, 0.0220 Ha, is subsequently applied. Radiation of this energy must be absorbed since the $1_e$ state is of a higher energy. Angular momentum conservation,  $^-2_e + 1_p$ = $^-1_e$, therefore predicts that the resulting excitonic state will have EAM = $^-1_e$, and that is exactly what was is found. 

The lower-right plot of Figure~\ref{ETCm2tom1} shows a completely different behavior for a sequence of pulses though. The first laser is identical to that in the figure at lower-left, and it produces an EAM = $^-2_e$. A second laser is then used that has the same energy as that used in the lower-left plot (0.0220 Ha) but with the opposite angular momentum, $^-1_p$. Once again, energy conservation demands that such radiation can only be absorbed and not emitted. However, the associated statement of AM conservation is now $^-2_e + (^-1_p)$ = $^-3_e$. This AM is not supported by the molecule, and the result is that radiation from the second laser cannot be absorbed. The lower-right plot shows that this is the case, where EAM = $^-2_e$ is maintained despite the application of the second laser--i.e. the molecule appears transparent to the second laser in this excitonic state.

As is clear in the top plot of Figure~\ref{ETCm2tom1}, the second laser needs to be much stronger than the first because the transition dipole between excited states is in the mid-arm region, where the relevant state densities are small, while the transition dipoles from the ground state are located in the arms where the relevant state densities are much larger. 

%
%

\begin{figure}[hptb]
\begin{center}
\includegraphics[width=0.9\textwidth]{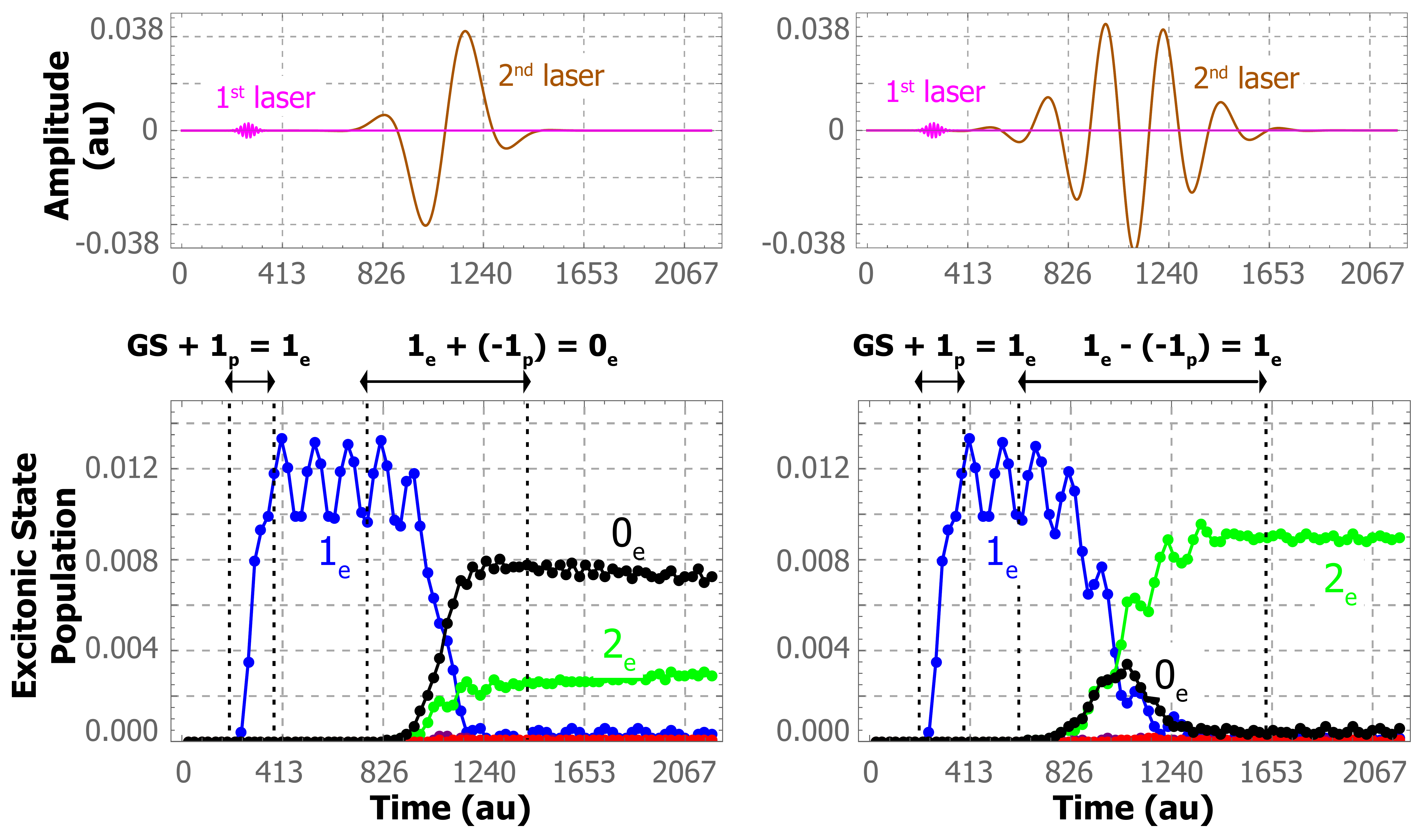}
\end{center}
\caption{\emph{TD-DFT AM manipulation:} $1_e + (^-1)_p = 0_e$ (left) and $1_e - (^-1)_p = 2_e$ (right).   Panel (a) shows two laser sequences with the associated population transfers given in panel (b).  In both processes, the parameters of Equation \ref{gaussian} for first laser pulses are $\{ t_0=272, \tau=27.2, \omega=0.390, E_0=0.00300 $. The parameters for second laser pulses are (left): $\{ t_0=1090, \tau=136, \omega=0.0158,  E_0=0.0500 \}$ and (right): $\{ t_0=1090, \tau=218, \omega=0.0257,  E_0=0.0500 \}$ in atomic units. }
\label{ETC1to2or0}
\end{figure}

A second comparative analysis, summarized in Figure \ref{ETC1to2or0}, shows how light with a fixed PAM can be adjusted in frequency to either increase or decrease EAM. In both scenarios shown, the molecule is first given an EAM = $1_e$ and, in both cases, a second laser pulse with PAM = $^-1_p$ is subsequently applied. In the evolution shown at lower-right, the energy of the second pulse is equal to the difference between the states $1_e$ and $0_e$, 0.0158 Ha (Table~\ref{CasidaExs}). This results in the absorption of radiation because the $0_e$ state is of higher energy with the AM balance equation of $1_e + (^-1)_p = 0_e$. On the other hand, tuning the second pulse to an energy of 0.0257 Ha causes the system to transition to the $2_e$ state because this is the energy difference between the $1_e$ and $2_e$ states (Table~\ref{CasidaExs}). In this case, radiation is emitted because the $2_e$ state is of lower energy, and the AM balance is $1_e -(^-1)_p = 2_e$.  

Note that there is an oscillation in the population of the $1_e$ state (blue curve) in both of the lower plots of Figure \ref{ETC1to2or0}. This is an artifact associated with the single-determinant approximation in concert with our piecewise homogeneous construction of the incident beam. This field approximation results in a larger contribution of the non-primary determinant, and the artificial oscillation in population is an indicator that the relative weighting of these determinants is time dependent. In contrast, the analogous curve associated with circularly polarized light (Figure~\ref{STOC}) shows no such oscillation because the non-primary determinants make almost no contribution. 

%
%

\begin{figure}[hptb]
\begin{center}
\includegraphics[width=1.0\textwidth]{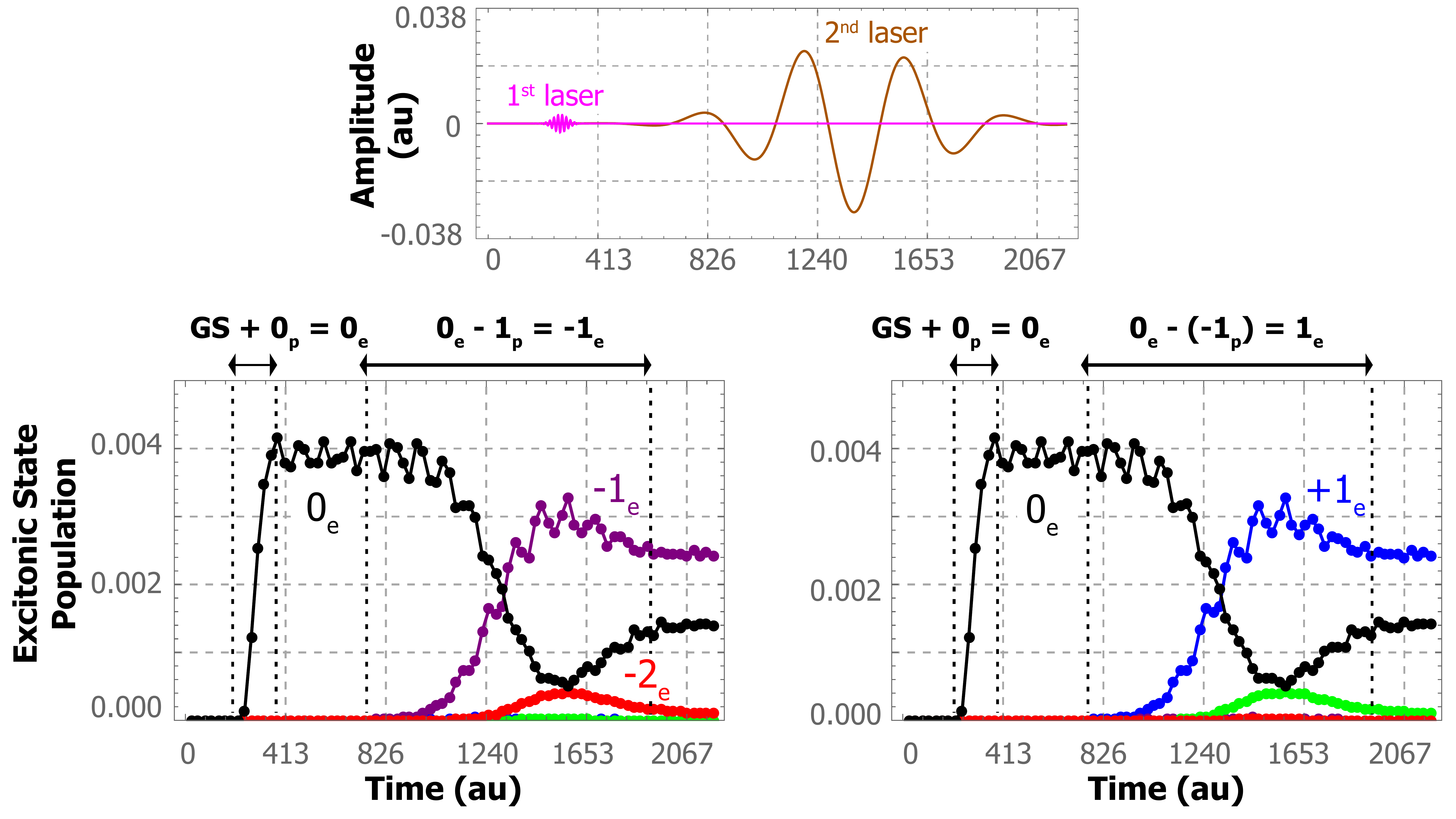}
\end{center}
\caption{\emph{TD-DFT PAM emissions:} $0_e - (1)_p$ = $^-1_e$ (left) and $0_e - (^-1)_p = 1_e$ (right).  Two cases are considered, and in both an initial laser pulse, shown in panel (a), generates a $0_e$ AM as shown in both plots of panel (b). The second laser pulse differs for each case though. A $^+1_p$ pulse stimulates emission and changes the excitonic state to $^-1_e$ (b, left), while a $^-1_p$ pulse also stimulates emission but changes the excitonic state to $^+1_e$ (b, right) . The parameters of Equation \ref{gaussian} for first laser are $\{ t_0=272, \tau=27.2, \omega=0.405, E_0=0.00300\}$ and those for the second laser are $\{ t_0=1360, \tau=272, \omega=0.0158, E_0=0.0300 \}$ in atomic units.}
\label{ETC0tom1orp1}
\end{figure}

A third comparative analysis, shown in Figure~\ref{ETC0tom1orp1}, demonstrates how EAMs of opposite sign can be created from the same initial state by changing both the laser frequency and the sign of its angular momentum. In both left and right scenarios, a molecule is placed in its highest energy state, $0_e$, after application of an appropriate laser pulse. A second laser with PAM = $^-1_p$ (lower-left) stimulates emission and changes the molecule to an EAM = $^-1_e$. The associated statement of AM conservation is  $0_e - (^+1)_p$ =$ ^-1_e$. On the other hand, illuminating the molecule with a PAM = $^-1_p$ (lower-right) also stimulates the emission of radiation but changes the system to an EAM = $^+1_e$. The associated statement of AM conservation is  $0_e - (^-1)_p$ = $^+1_e$. 

%
%

\begin{figure}[hptb]
\begin{center}
\includegraphics[width=0.7\textwidth]{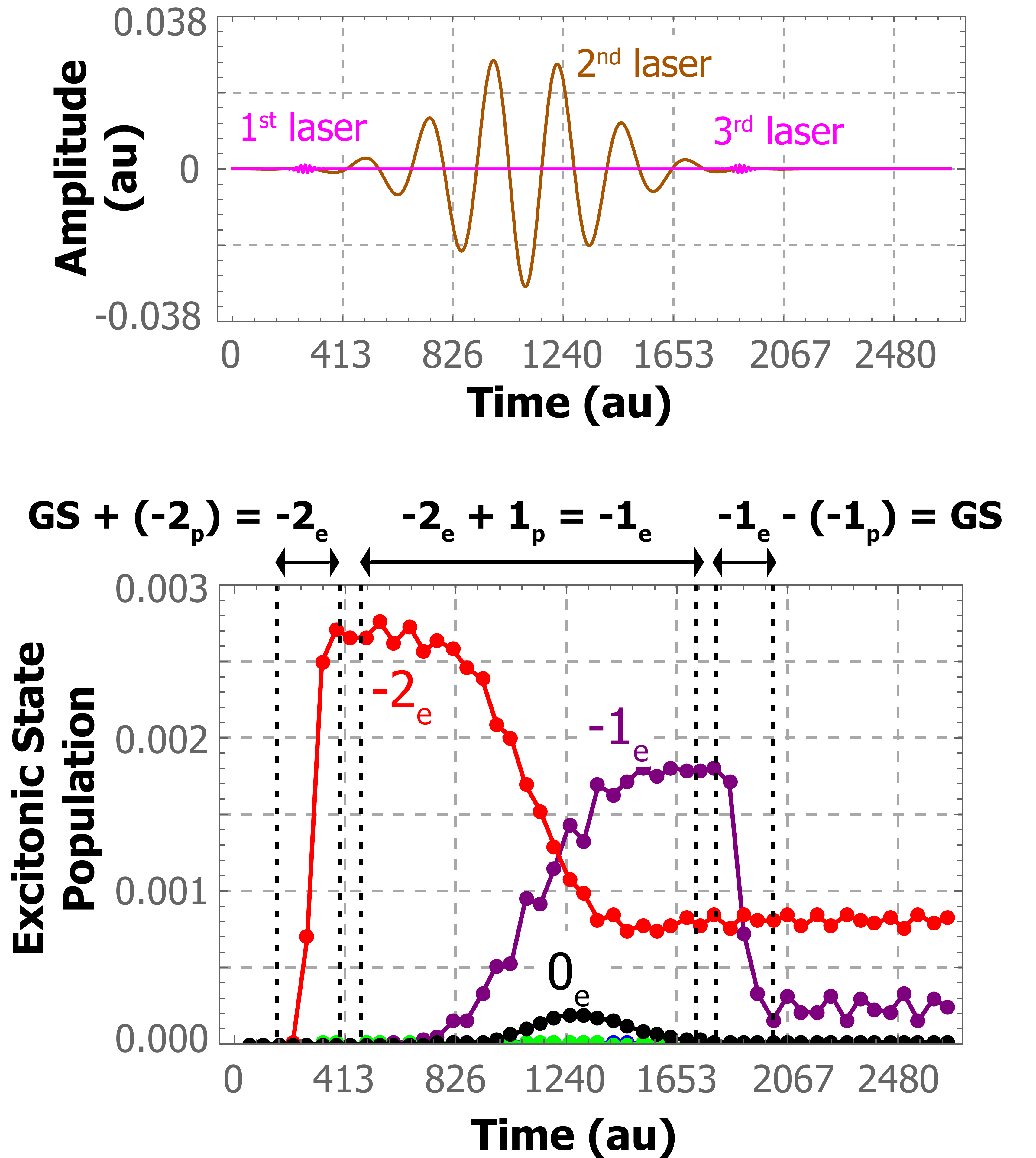}
\end{center}
\caption{\emph{TD-DFT PAM conversion:}  $^-2_p + 1_p$ = $^-1_p$. Panel (a) shows two laser pulses.  Panel (b) shows the system is excited into $^-2_e$ state by the first laser. Subsequent absorption of a $1_p$ radiation then transfers the system into $^-1_e$ state. This EAM is converted to PAM with a third laser which brings system back to its ground state by stimulation emission. The laser parameters of Equation \ref{gaussian} are: (first) $\{ t_0=272, \tau=27.2, \omega=0.367, E_0=0.00100 \}$; (second) $\{ t_0=1090, \tau=272, \omega=0.0257, E_0=0.0300 \}$; and (third)$\{t_0=1905, \tau=27.2, \omega=0.390, E_0=0.00100\}$ in atomic units. }
\label{ETCemission}
\end{figure}

All three of these comparative TD-DFT analyses have focused on EAM arithmetic. When purposed as a PAM converter, though, it is important to be able to transfer the final AM into an electromagnetic field. Such an operation, already considered within the TB setting, can be demonstrated in a more realistic TD-DFT simulation. Figure~\ref{ETCemission} summarizes a simulation in which the molecule is first given a $^-2_e$ AM which is subsequently changed to $^-1_e$ using a second laser. Taken together, these first two steps  can be viewed as PAM addition: $^-2_p + 1_p$ = $^-1_e$. A third laser then stimulates the emission of PAM = $^-1_p$ leaving the system in its ground state. The photonic arithmetic for this conversion process is therefore  $^-2_p + 1_p$ = $^-1_p$.

%
%

The excited state excitonic populations are small because very weak and short lasers are applied in all TD-DFT simulations. This is a pragmatic step taken to avoid the population of extraneous eigenstates that result from stronger or longer laser pulses. This is purely a computational issue that results because the vector vortex beams were approximated with five piecewise-homogeneous components---a computational work-around to the limitations of the input fields allowed by the OCTOPUS TD-DFT software. The result is unphysical light-matter interactions in the regions between each arm that can be remedied by dividing the region into more than five piecewise-homogeneous components. Then stronger and/or longer laser illumination can be applied to increase the population of twisted excitons.

\subsection{Algebra of Circularly Polarized Lights}

As an alternative to applying an radial vector vortex, PAM can be inputted using circularly polarized light. This just amounts to a change of basis for describing the molecular dipoles since circularly polarized light can be mathematically decomposed into a combination of radial and azimuthal vector vortices\cite{CircPolVortexOL2006}:
\begin{equation}
\frac{1}{\sqrt{2}}(\vec{e}_x\pm\imath\vec{e}_y)=\frac{1}{\sqrt{2}}\mathrm{e}^{\pm \imath \phi}(\vec{e}_r\pm\imath\vec{e}_\phi). \label{spinlight}
\end{equation}
Here $\{\vec{e}_x, \vec{e}_y\}$ and $\{\vec{e}_r, \vec{e}_\phi\}$ are the basis vectors in Cartesian and polar representations. In the 5-arm $H_2$ system, only the radial vortex components are absorbed. Through a series of absorption events, circularly polarized light can be used to generate vector vortices with an arbitrary AM.

\begin{figure}[hptb]
\begin{center}
\includegraphics[width=0.7\textwidth]{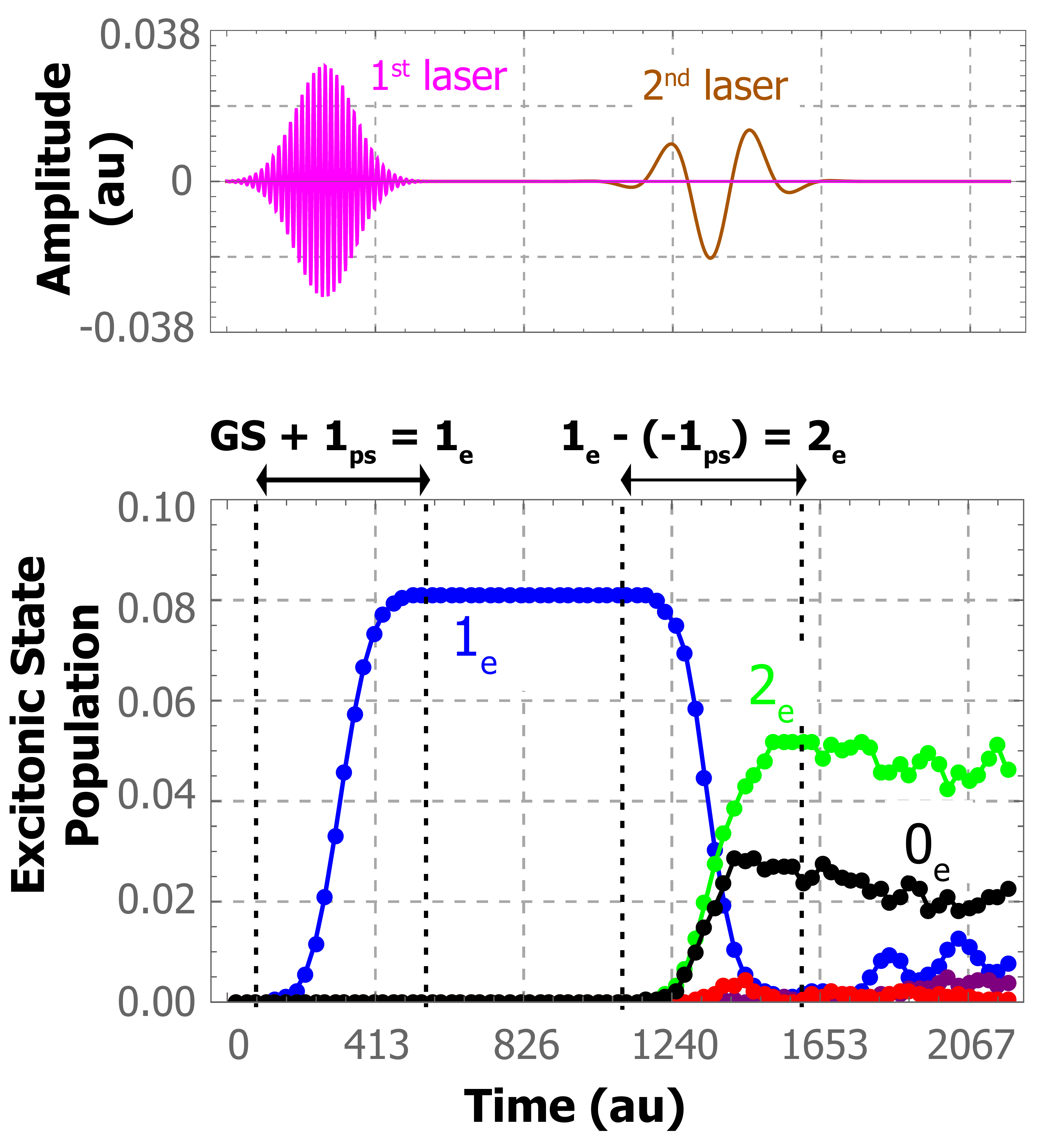}
\end{center}
\caption{\emph{Addition of circularly polarized light:} $1_{ps} - (^-1)_{ps} = 2_e$. Panel (a) shows two laser pulse.  Panel (b) shows the system is sequentially provided with two units of AM resulting in $2_e$.  Subsequent emission would produce $2_{p}$ radiation. The laser parameters are: (first) $ \{t_0=272, \tau=81.5, \omega=0.390, E_0=0.00300 \}$; and (second) $\{ t_0=1360, \tau=109, \omega=0.0257, E_0=0.00200 \}$ in atomic units.}\label{STOC}
\end{figure}

This is demonstrated in Figure \ref{STOC}, where a sequence of circularly polarized laser pulses are applied. The first laser pulse causes absorption and the following AM conservation relation: $GS + 1_{p} = 1_e$. The second laser, of opposite spin, results in emission: $1_e -(^-1)_{p} = 2_e$.  

%
%

Figure~\ref{STOC} shows that the strength of the first laser is the same as those used for the vortex beams of Figure~\ref{ETCm2tom1}, \ref{ETC1to2or0}, and \ref{ETC0tom1orp1}, but the duration is three times longer. This allows a much larger population of proper excited states without involvement of extraneous eigenstates. As compared with Figure~\ref{ETC1to2or0}, we now have a very smooth population of the $1_e$ state (blue curve). This confirms that the origin of the oscillations in our former TD-DFT simulations is actually the approximation of the vector vortex with five piecewise-homogeneous components.

\section{Conclusions}\label{conclude}
Light and molecules can be engineered so that quanta of angular momentum can be exchanged between the two. Strictly speaking, the transformation is  between the AM of a radiation field (PAM) and a quasi-angular momentum (EAM). This makes it possible to design processes in which sequential laser pulses are used to increase or decrease the EAM. Subsequent emission results in radiation with a different PAM than the input light. Unlike existing approaches, this molecular strategy offers a means of manipulating the angular momentum of light which does not rely on the nonlinear optical properties of a mediating crystal. Computational proofs of concept were provided using tight-binding theory as well as the more realistic many-body setting of time-domain density functional theory. While the present work focused on stimulated emissions, AM conversions may culminate in spontaneous emission as well. 

Angular momentum conservation, valid within a paraxial approximation, and algebraic manipulations were elucidated using simple tight-binding Hamiltonians. However, time-domain DFT gives consistent results within a many-electron setting with the effects of electron correlation and exchange accounted for. In both settings, the important influence of phonon entanglement and dynamic disorder have been neglected. It was assumed that phase coherence is maintained in the superposition of arm excitons that comprise molecular eigenstates. Unless the electronic coupling between arms is sufficiently strong to preserve coherence in the face of these effects, they will set a time scale over which AM manipulations must be carried out for a given temperature~\cite{Ishizaki_2009}. Moreover, AM conversions must compete favorably with energy relaxation pathways such as fluorescence and internal conversion~\cite{LaCount_2017}.

Since changes to the excitonic angular momentum are accompanied by a change in energy---e.g Table \ref{TB_data}---these AM manipulations can be viewed as a special type of laser-based upconversion or downconversion that can even be carried out with either pulsed or continuous-wave lasers. A number of upconversion methodologies have been experimentally realized, for instance, and these offer a path forward for manipulating the angular momentum of light~\cite{Scheps_1996}. Laser-based energy conversion methodologies tend to use solid-state crystals, so the approach would need to be adapted to molecular media.  The sequential banking of AM requires that conversions must be fast relative to competing molecular relaxation processes. The entire AM conversion process of Figure \ref{ETCemission} takes only 44 fs, and this is orders of magnitude faster than the nanosecond time scales for internal conversion and photoluminescence for typical organic molecules~\cite{LaCount_2017}.  The issues to be faced here are also encountered in a range of energy upconversion and downconversion strategies~\cite{Chen_2014}. 

The level of tight-binding analysis employed does not require any details of the molecular structure beyond point group, and the time-domain DFT analyses adopted, as a computational expedient, molecular arms composed of hydrogen dimers. There exist a panoply of molecules which exhibit $C_N$ or $C_{Nh}$ symmetry, though, with only a few  examples shown in Figure \ref{molecules}. Polycyclic aromatic hydrocarbons may be promising candidates of this type, and several examples are shown in the figure. Their aromaticity counters the effects of dephasing due to vibrations. Another intriguing possibility is to functionalize inert scaffolds that have the requisite point symmetry. 

\begin{figure}[hptb]
\begin{center}
\includegraphics[width=0.7\textwidth]{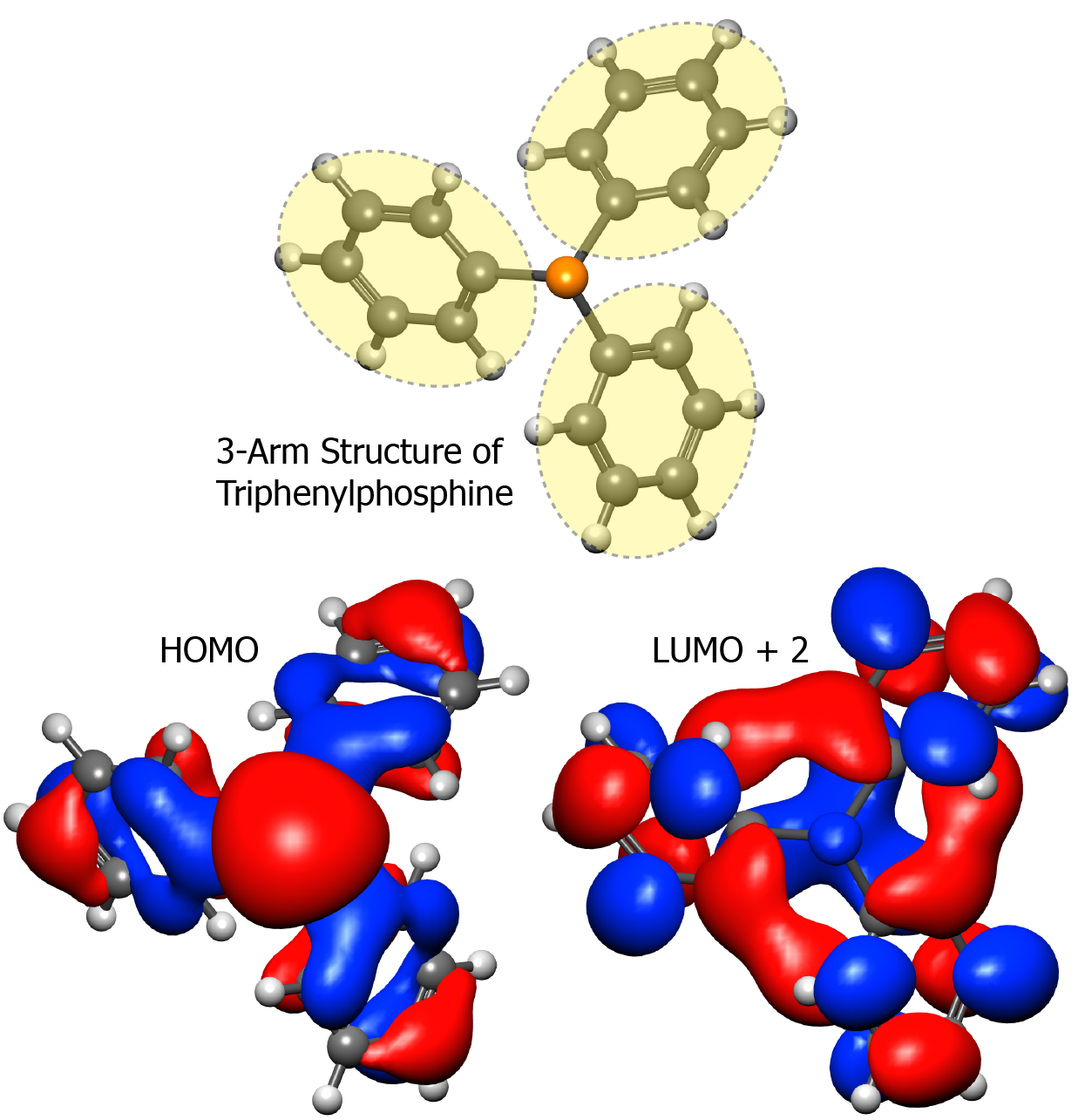}
\end{center}
\caption{\emph{High-energy exciton of $\rm{Ph}_3\rm{P}$ with EAM = $0_e$.} The $C_3$ symmetry of $\rm{Ph}_3\rm{P}$  (a) results in an exciton composed of HOMO and LUMO+2 orbitals that also has $C_3$ symmetry and so an EAM = $0_e$. The exciton energy is 3.47 eV. The red and blue isosurfaces in (b) are for densities for 0.02 ${\rm bohr}^{-3/2}$ of the real part of wave functions. Colors correspond to plus (red) and minus (blue) values.} 
\label{TPP_1}
\end{figure}

\begin{figure}[hptb]
\begin{center}
\includegraphics[width=0.7\textwidth]{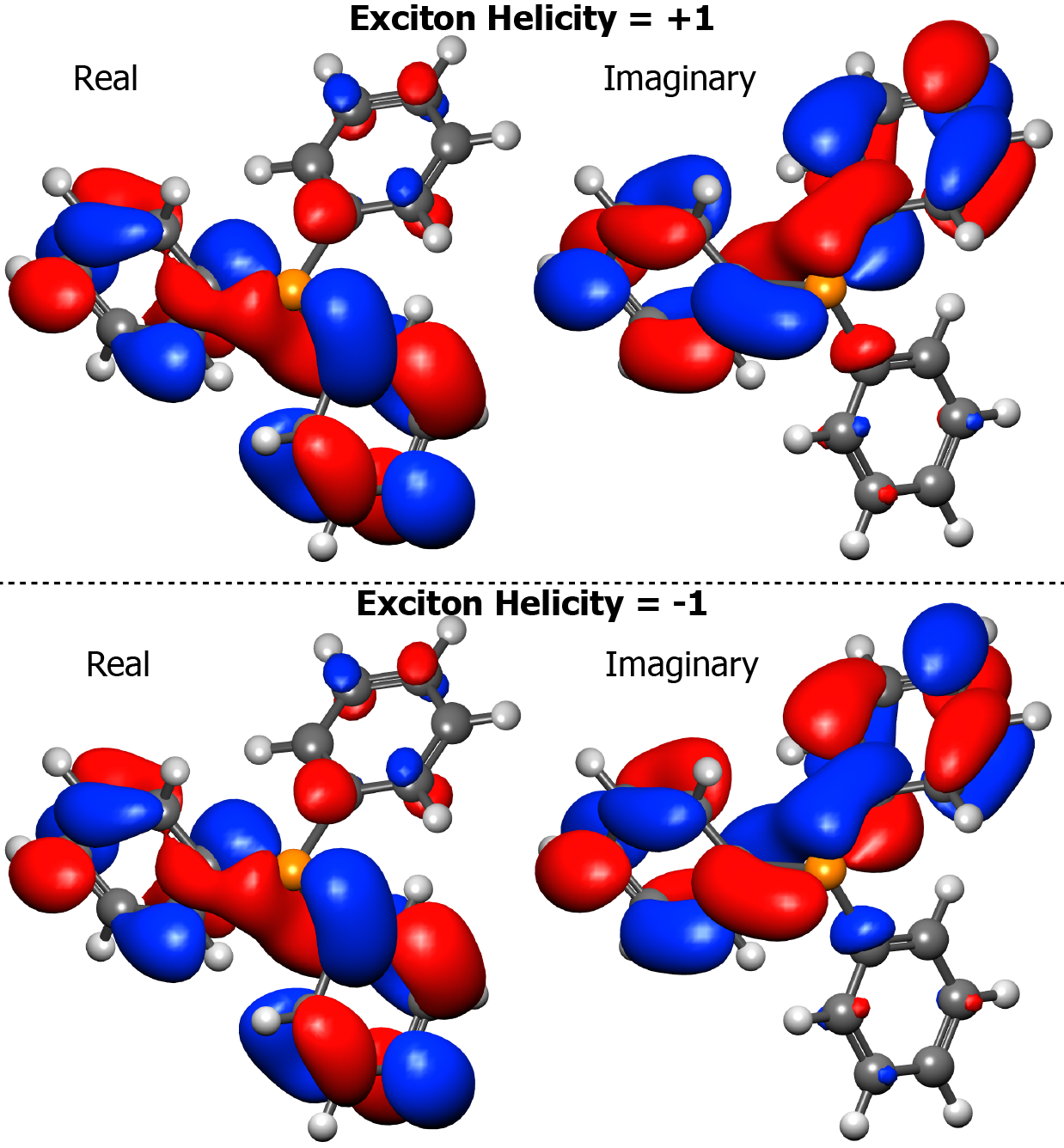}
\end{center}
\caption{\emph{Low-energy excitons of $\rm{Ph}_3\rm{P}$ with EAM = $^\pm 1_e$.} Orbitals with opposite 120-degree phase progressions can be constructed from the LUMO and LUMO+1 orbitals of $\rm{Ph}_3\rm{P}$. Panel (a) shows the orbital for EAM = $^+1_e$ and panel (b) shows the orbital for EAM = $^-1_e$. Their energies are each 3.43 eV, lower than the exciton with no angular momentum. The red and blue isosurfaces of wave function densities are for 0.02 ${\rm bohr}^{-3/2}$. Colors correspond to plus (red) and minus (blue) values. The real parts of the two orbitals are identical but the imaginary parts are of opposite sign. This is consistent with the opposing 120-degree phase progressions.} 
\label{TPP_2}
\end{figure}

Screening of candidate molecules can be carried out using simple DFT analysis to identify the requisite excitonic structures.  As an example, the DFT orbitals of $\rm{Ph}_3\rm{P}$ can be used to generate a rudimentary estimate the electronic structures associated with its three EAM states: +1, -1 and 0. The EAM = $0_e$ state is composed of the highest occupied molecule orbital (HOMO) and the third excited state--i.e. two states above the lowest unoccupied molecular orbital (LUMO). These both have the requisite $C_3$ symmetry as shown in Figure \ref{TPP_1}. The associated exciton energy is 3.47 eV (357 nm). On the other hand, the LUMO and LUMO+1 are degenerate and can be combined to create states with EAM = $^\pm 1_e$ as shown in Figure \ref{TPP_2}. Taken with the HOMO, the associated excitons have an energy of 3.43 eV (361 nm), lower than the EAM = $0_e$ state and consistent with both the TB results of Table  \ref{TB_data} and the $H_2$ trimers considered with TD-DFT. These states are constructed by first adding the LUMO and LUMO+1 orbitals, then assigning orbitals to each arm by localizing this sum as shown in the top panel of Figure \ref{TPP_1}. These arm orbitals are then given a 120-degree phase progression to obtain the structures of Figure \ref{TPP_2}. This progression is evident in the figure, where it is clear that the two orbitals have the same real part and imaginary parts of opposite sign. As a check, the sum of the squares of the projections of these states with the original LUMO and LUMO+1 orbitals was found to be 0.98 in each case. While carried out in a very crude way, with simple combinations of DFT orbitals to represent excitons, this analysis serves to demonstrate that light at the far edge of the visible spectrum can be used to create twisted excitons.

%
%
%

%
%


\section{Methods}\label{TD-DFT}

Real-time simulations are made possible through the Runge-Gross (RG) reformulation of time-dependent Schr{\" o}dinger  equation~\cite{RGtddftPRL1984}:
\begin{equation}
\mathrm{i}\frac{\partial}{\partial t}\psi_i(\vec r,t) = \Big[ -\frac{1}{2}\Delta^2 +\nu_{Ha}[\rho](\vec r,t) +\nu_{xc}[\rho](\vec r,t)+\nu_{ext}(\vec r,t)+\nu_{lm}(\vec r,t)\Big]\psi_i(\vec r,t).
\label{RG}
\end{equation}
Here $\nu_{Ha}$ and $\nu_{xc}$ are the Hartree and exchange-correlation potentials, respectively, and $\nu_{ext}$ is the external potential representing all nuclei. The semi-classical light-matter interaction term, $\nu_{lm}$, has the form $\vec R\cdot\vec{E}$ within electronic dipole approximation adopted here, where $\vec R:=\sum_i\vec{r}_i$ is the Kohn-Sham position operator and atomic units are used. The spin-reduced electronic density, $\rho(\vec r, t)$, is expressed in terms of thes time-dependent Kohn-Sham (TDKS) orbitals, $\psi_i(\vec r, t)$, as
\begin{equation}
\rho(r,t)=\sum_i^{N}|\psi_i(r,t)|^2.
\label{dens}
\end{equation}
These orbitals, in turn, can be represented in the basis of their counterparts at time zero, $\psi_i$, so that the time-propagated multi-electron wavefunction is constructed from a linear combination of determinants built from these initial orbitals\cite{TimePropState}:
\begin{equation}
\ket{\Psi(t)} = c_0(t)\ket{\Psi_{gs}} + \sum_a^{occ}\sum_i^{unocc}c_a^i(t)\ket{\Psi_a^i} .
\label{TPstate}
\end{equation}
Ket $\ket{\Psi_{gs}} = \ket{\psi_1\psi_2\psi_3\psi_4\psi_5}$ is the ground state and $\ket{\Psi_a^i} = \ket{\psi_1\cdots\psi_i\cdots\psi_5}$ is a determinant with the $a^{th}$ occupied Kohn-Sham (KS) orbital replaced by the $i^{th}$ unoccupied orbital. The first summation is over all occupied KS orbitals, five occupied KS orbitals in the case of 5-arm $H_2$ system, and the second summation is over all unoccupied KS orbitals. In our case, because the frequency of laser is chosen to only access the first five lowest excited states, the only unoccupied KS orbital is the sixth as shown in Table~\ref{CasidaExs}. 

If it was possible to express Equation~\ref{TPstate} in form of Equation~\ref{TBstates}, the associated EAM could be determined directly. Such a simple expansion of $\ket{\Psi(t)}$ in basis of $\ket{e_j}$ does not exist, though, since it is a many-body wavefunction. This is remedied, albeit in an approximate way, by working only with the dominant determinant for which only lowest unoccupied molecular orbital (LUMO) is involved in Equation~\ref{TPstate}. This makes it possible to combine the determinants corresponding to each EAM subspace  $\{^\pm m\}$ with $m=0,1,2\cdots$, as in Equation~\ref{Rearranged}, allowing the EAM of $\ket{\Psi(t)}$ to be obtained via KS orbitals using the method detailed below. 

In all simulations, twisted excitons are a constructed as a linear combination of corresponding pairs of degenerate excited states, consistent with the TB model. Focusing on the 5-arm system, if a laser with PAM = $0_p$ is applied, then the resulting excited state can be approximated with a determinant involving only $\Psi_1^6$. Likewise, the application of PAM = $^\pm1_p$ results in excited states that can be approximated with determinants involving only $\Psi_2^6$ and $\Psi_3^6$,  and PAM = $^\pm 2_p$ yields states that are well-approximated with only $\Psi_4^6$ and $\Psi_5^6$. The time-propagated wavefunction for each EAM subspaces $\{0\}$, $\{^\pm1\}$ and $\{^\pm2\}$ can therefore be expressed, respectively, as:
\begin{eqnarray}
c_1^6(t)\ket{\Psi_1^6}&=&\ket{c_1^6(t)\psi_6,\psi_2\psi_3\psi_4\psi_5}\label{Rearranged} \\ 
c_2^6(t)\ket{\Psi_2^6} + c_3^6(t)\ket{\Psi_3^6} &=&\ket{\psi_1,c_3^6(t)\psi_2-c_2^6(t)\psi_3,\psi_6\psi_4\psi_5}\nonumber\\
c_4^6(t)\ket{\Psi_4^6} + c_5^6(t)\ket{\Psi_5^6} &=& \ket{\psi_1\psi_2\psi_3,c_5^6(t)\psi_4-c_4^6(t) \psi_5,\psi_6} .\nonumber
\end{eqnarray}
%

%
%
\begin{figure}[hptb]
\begin{center}
\includegraphics[width=0.8\textwidth]{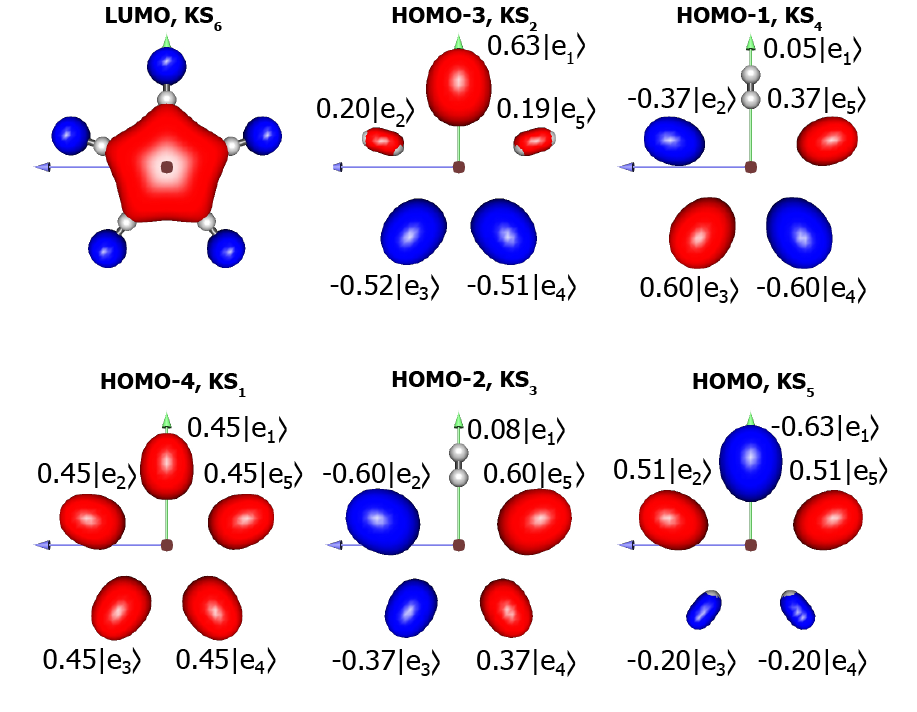}
\end{center}
\caption{\emph{Decomposition of KS orbitals of 5-arm $H_2$ system.} The five occupied KS orbitals have been expressed in the basis of arm wavefunctions, $\ket{e_j}$, with their coefficients labeled on the corresponding arms. The red (blue) isosurfaces indicate positive (negative) values of the wavefunctions. The LUMO is given in order to show that it is symmetric. HOMO = highest occupied molecular orbital. }\label{KSorbitals}
\end{figure}

The only difference among these three equations is that the ground state determinant is modified as follows: $\psi_1$ replaced by $c_1^6(t)\psi_6$; $\psi_2$ and $\psi_3$ are replaced by $c_3^6(t)\psi_2-c_2^6(t)\psi_3$; and $\psi_6$, $\psi_4$ and $\psi_5$ are replaced by $c_5^6(t)\psi_4-c_4^6(t)\psi_5$ and $\psi_6$. The ground state determinant $\ket{\psi_1\psi_2\psi_3\psi_4\psi_5}$ is the $0_e$ state of course. These replacement orbitals must therefore be responsible for the EAM of excited states. Figure~\ref{KSorbitals} gives the isosurface and decomposition in the basis of $e_j$ with $j\in\{1,\cdots,5\}$ of all the relevant KS orbitals. As shown in Figure~\ref{KSorbitals}, the LUMO is symmetrically distributed across all five arms. Therefore $\psi_6$ will not introduce a phase difference among arms in the right side of Equation~\ref{Rearranged}. This implies that  $c_3^6(t)\psi_2-c_2^6(t)\psi_3$ and $c_5^6(t)\psi_4-c_4^6(t)\psi_5$ will introduce a phase dependence corresponding to $^\pm1_e$ and $^\pm2_e$, respectively. The population of each twisted exciton state is therefore given by:
\begin{eqnarray}
P_{0_e}&=&2|c_1^6(t)|^2\nonumber\\
P_{^-1_e}&=&2|\braket{v_{^-1}|c_3^6(t)\psi_2-c_2^6(t)\psi_3}|^2\nonumber\\
P_{1_e}&=&2|\braket{v_{1}|c_3^6(t)\psi_2-c_2^6(t)\psi_3}|^2\nonumber\\
P_{-2_e}&=&2|\braket{v_{^-2}|c_5^6(t)\psi_4-c_4^6(t)\psi_5}|^2\nonumber\\
P_{2_e}&=&2|\braket{v_{2}|c_5^6(t)\psi_4-c_4^6(t)\psi_5}|^2.
\label{population}
\end{eqnarray}
Here $\ket{v_{q_e}}$ is the eigenstate associated with an EAM of $q_e$, from Equation~\ref{TBstates}, and the factor of two in each expression accounts for the fact that the electron spin can be either up or down.

\section{Acknowledgements}
We are grateful to Profs. David L. Andrews, Mark E. Siemens and Guillermo F. Quinteiro for extended discussions on the generation of twisted light. All calculations were carried out using the high performance computing resources provided by the Golden Energy Computing Organization at the Colorado School of Mines.


\end{document}